\documentclass[%
reprint,
amsmath,amssymb,
aps,showkeys,
pre
]{revtex4-1}

\usepackage{graphicx}
\usepackage{amsmath,bm}
\usepackage{amsmath}
\usepackage{amssymb}
\usepackage{graphicx}\usepackage{epsfig}
\usepackage{setspace}
\usepackage{color}
\usepackage{float}

\newcommand{\be}{\begin{equation}}
\newcommand{\ee}{\end{equation}}

\newcommand{\bea}{\begin{eqnarray}}
\newcommand{\eea}{\end{eqnarray}}

\renewcommand{\Im}{{\rm \, Im\,}}

\definecolor{nvgreen}{rgb}{0.0, 0.5, 0.2}

\newcommand{\addGF}[1]{\textcolor{black}{#1}}

\begin{document}
\sloppy

\title{Fibonacci turbulence}
\author{ Natalia Vladimirova$^{1,2}$, Michal Shavit$^{1}$ and Gregory Falkovich$^{1}$}
\affiliation{Weizmann Institute of Science, Rehovot 76100 Israel\\
 Brown University,  Providence, RI 02912, USA}

\date{\today}

 \begin{abstract}
Never is the difference between thermal equilibrium and turbulence so dramatic, as when a quadratic invariant  makes the equilibrium statistics exactly Gaussian with independently fluctuating modes. That happens in two very different yet deeply connected classes of systems: incompressible hydrodynamics and resonantly interacting waves. This work presents the first detailed information-theoretic analysis of turbulence in such strongly interacting systems. The analysis  {involves both energy and entropy} and elucidates the fundamental roles of space and time in setting the cascade direction and the changes of the statistics along it. We introduce a beautifully simple yet rich family of discrete models with triplet interactions  {of neighboring modes} and show that it has  quadratic conservation laws defined by the Fibonacci numbers.  {Depending on how the interaction time changes with the mode number}, three types of turbulence were found: single direct cascade, double cascade, and the first ever case of a single inverse cascade.  We describe  quantitatively how deviation from thermal equilibrium all the way to turbulent cascades makes statistics increasingly non-Gaussian and find the self-similar form of the one-mode probability distribution. We reveal where the information (entropy deficit) is encoded and disentangle the communication channels  between modes, as quantified by the mutual information in pairs and the interaction information inside triplets.
\end{abstract}

\maketitle


\section{Introduction}
Existence of quadratic invariants and Gaussianity of equilibrium in a strongly interacting system may seem exceptional. {Indeed, generic systems have no invariants except Hamiltonian. Strongly interacting systems have non-quadratic Hamiltonians, so that equilibrium Gibbs distribution (the exponent of the Hamiltonian) is generally non-Gaussian.  And yet two very distinct wide classes of physical systems have quadratic invariants and Gaussian statistics at thermal equilibrium.}  The first class is the family of hydrodynamic models, starting from the celebrated hydrodynamic Euler equation and including many equations for geophysical, astrophysical and magnetohydrodynamic flows. The second class, as will be described in this paper, contains systems of resonantly interacting waves. We show that the discretized models of the first class exactly correspond to the second one. We shall consider one particular (arguably the simplest) family of such models and describe far-from equilibrium (turbulent) states of such systems.

 {One calls turbulence a state of any system, where many degrees of freedom are deviated far from thermal equilibrium. Therefore, studies of turbulence encompass a wide variety of phenomena in nature and industry, from pipe flows to ripples on a paddle. It can be studied from the viewpoint of a mathematician, engineer or a physicist. Here we employ the perspective of statistical physics, which is interested in fundamental principles that determine statistical distributions in turbulence and thermal equilibrium.  We shall use both the traditional viewpoint of cascades and the relatively recent viewpoint of information theory, that is we address both energy and entropy of turbulence. So far, statistical physics approach to turbulence was to a large extent  devoted to two quite distinct classes:  systems of interacting waves like those on the surface of the ocean or a paddle and incompressible vortical flows where no waves are possible. Here we build a bridge between these two classes and show that discrete models of a certain kind can describe both.}

 {On the one hand, the vorticity, ${\bf \omega}=\nabla\times{\bf v}$, of an isentropic flow of incompressible fluid satisfies the Euler equation: $ \partial{\bf \omega}/\partial t=\nabla\times({\bf v}\times {\bf \omega})$.} Quite similar are two-dimensional hydrodynamic models, where a scalar field $a$ (vorticity, temperature, potential) is linearly related to the stream function $\psi$ of the velocity carrying the field: $\partial a/\partial t=-({\bf v}\cdot\nabla)a$, ${\bf v}=(\partial\psi/\partial y,-\partial\psi/\partial x)$, $\psi({\bf r})=\int d{\bf r}'|{\bf r}-{\bf r}'|^{m-2}a({\bf r}')$. For the 2D Euler equation, $m=2$. Other cases include surface geostrophic ($m=1$),  rotating shallow fluid or magnetized plasma ($m=-2$), etc. After Fourier transform,
\be {\dot a_{\bf k}}=\sum\nolimits_{\bf q}\![{{\bf k}\times{\bf q}}]q^{-m}a_{\bf q}a_{\bf k-q}\,.\label{M}\ee
 {All such equations  have quadratic nonlinearity and  quadratic invariants. Then it was} suggested \cite{Obukhov} to model different cases of fluid turbulence by the chains of ODEs having quadratic invariant  $g_{ij}u^iu^j$ and these properties: \be   \dot u_i=\Gamma^i_{jl}u_ju_l\,,\quad \Gamma^i_{il}=0= g_{ik}\Gamma^k_{jl}+ g_{lk}\Gamma^k_{ji}+ g_{jk}\Gamma^k_{li}\,.\label{Ob}\ee

 {On the other hand, consider} resonantly interacting waves with the general Hamiltonian,
\be{\cal H}_w =\sum\nolimits_i\omega_i  |b_i|^2  +  \sum\nolimits_{ijl} \left(V_{l, ij}  b_i^{*}b_j^{*}b_l+  V_{l, ij} ^*b_ib_jb_l^{*}\right)\,,\label{HamW}\ee
where $V_{l, ij} \not=0$ only if $\omega_i+\omega_j=\omega_l$. By the gauge transformation, $a_i=b_i\exp(\imath \omega_i t)$,
we can turn the equations of motion, $\imath \dot b_i={\partial{\cal H}_w/ \partial b_i^*}$ into a  system of the type  (\ref{M},\ref{Ob}):
\be\imath \dot a_i =\sum\nolimits_{jl}\bigl(V_{i,jl}^*a_j a_l+2V_{l, ij} a_j^{*}a_l \bigr)\,.\label{Ham1}\ee
This means that quadratic and cubic parts of the Hamiltonian are conserved separately. If such a system is brought into contact with thermostat, it is straightforward to show that the statistics is Gaussian: $\ln {\cal P}\{a_i\}\propto     -\sum\nolimits_i\omega_i  |a_i|^2$. 

Our interest in resonances is connected to that in non-equilibrium. Thermal equilibrium does not distinguish between resonant and non-resonant interactions because of the detailed balance: whatever correlations can be built over time between resonantly interacting modes, the reverse process destroying these correlations is equally probable. This is not so away from thermal equilibrium, especially in turbulence.

Neglecting non-resonant and accounting only resonant interactions is the standard approach to weakly interacting systems, even though the weak nonlinearity assumption breaks for resonant modes. Weak turbulence theory gets around this by considering continuous distribution and integrating over resonances to get  the  kinetic wave equation,  {which describes nonlinear evolution that is slow compared to linear oscillations with wave frequencies}  \cite{Peierls,ZLF,NR,Naz}. There is a  {tendency} in theoretical statistical physics to restrict consideration to two opposite limits: either treat few modes or infinitely many. That {preference} is even stronger in the studies of non-equilibrium. And yet not only most of the real-world phenomena fall in between these limits, but, as we show here, one learns some fundamental lessons comparing equilibrium and non-equilibrium states of systems with a finite number of degrees of freedom, where phase coherence can play a prominent role. A similar lesson condensed matter physics taught us by discovering the world of mesoscopic phenomena,  {where the system size was made smaller than the phase coherence length}.

The previous treatment of mode discreteness was focused on the sparseness of resonances for the particular cases when resonant surfaces  {$\omega_k+\omega_q=\omega_{|\bf k+q|}$} did not pass through integer lattice  {determined by a box} \cite{Naz,Kart}. Yet in many cases resonance surfaces lay in the lattice. For example, in a quite generic case of quadratic dispersion relation, $\omega_k\propto k^2$, Pythagorean theorem makes the resonance surface for three-wave interactions just perpendicular to any wavevector, so that in any rectangular box resonantly interacting triads fill the lattice of  {the box eigen} modes.

Class of models (\ref{M},\ref{Ob},\ref{Ham1}) is ideally suited for the comparative analysis of thermal equilibrium and turbulence. We show here that such analysis sheds light on the most fundamental aspects of turbulence, particularly the roles of spatial and temporal scales in determining cascade directions and build-up of intermittency.   We consider the particular sub-class of models that allow only neighboring interactions, and find it the most versatile tool to date to study turbulence as an ultimate far-from-equilibrium state. We carry here such detailed study of the known types of direct-only and double cascades with unprecedented numerical resolution. Even more important, our models allow for an inverse-only cascade never encountered before.

\section{Fibonacci turbulence}

We consider a sub-class of the models (\ref{M},\ref{Ob},\ref{Ham1}) which is Hamiltonian with a local interaction:
\begin{align}
     {\cal H} =\sum\nolimits_{i}  V_i\left( a_i^{*}a_{i+1}^{*}a_{i+2}+  a_ia_{i+1}a_{i+2}^{*}\right).\label{FibA}
 \end{align}
 The equations of motion $\imath  \dot a_i= {\partial{\cal H}/\partial a_i^*}$ are as follows:
 \begin{align}
\imath  \dot a_i&
=  V_{i-2}a_{i-1} a_{i-2}+V_{i-1}a_{i-1}^{*}a_{i+1}+V_ia_{i+1}^{*}a_{i+2}.\label{Fib}
 \end{align}
This \addGF{family of models (each characterized by $V_i$)} can have numerous classical and quantum applications, since $i$ can be denoting  real-space sites, spectral modes, masses of  particles, number of monomers in a polymers, etc. The Hamiltonian describes, in particular, decay and coalescence of waves or quantum particles, breakdown and coagulation of particles or polymerization  of polymers, etc, when interactions of comparable entities are dominant.
In particular, the model describes the resonant interaction of waves whose frequencies are the Fibonacci numbers $F_i=\{1,1,2,3,5\ldots\}$ defined by the identity $F_i+F_{i+1}=F_{i+2}$ with $F_0=0$. Indeed, such waves are described by the  Hamiltonian
\begin{align}\!\!\! {\cal H}_0= \sum\nolimits_{i}\! \left[F_i|a_i|^2+ V_i\left( a_i^{*}a_{i+1}^{*}a_{i+2}+  a_ia_{i+1}a_{i+2}^{*}\right)\right]. \label{Fib00}\end{align}
The first term corresponds to the linear terms in the equations of motion, while the second term   represents the only possible resonant interactions, since no non-consecutive Fibonacci numbers sum into another Fibonacci number (Zeckendorf theorem).
For any real $t$, the Hamiltonian (\ref{Fib00}) is invariant under the $U(1) \times U(1)$  transformation $a_i\to a_ie^{\imath F_it}$ due to  $F_i+F_{i+1}=F_{i+2}$.
The transformation  (to the wave envelopes) reduces the equation of motion  $\dot a_i= \partial{\cal H}_0/\partial a_i^* $ to  (\ref{Fib}).

If $i$ are spectral parameters, they are usually understood as shell numbers. \addGF{That means that one can {\it define} wave numbers as $k=F_i=[\phi^i-(-\phi)^{-i}]/\sqrt5$, where $\phi=(1+\sqrt5)/2$ is the golden mean. It plays here the role of an intershell ratio, since asymptotically at $|i|\gg1$, the wave number depends exponentially on the mode number:  $F_i\propto \phi^{|i|}$.} The model (\ref{Fib}) thus belongs to the class of the so-called shell models \cite{Bif}, that is (\ref{Ob}) with neighboring interactions. Coefficients of shell models are chosen to have one or two quadratic integrals of motion.
In particular, the  Sabra shell model \cite{Pro,Lvo} for a particular choice of coefficients (non-surprisingly, connected by the golden ratio)   coincides with (\ref{Fib}), which is Hamiltonian and has the cubic integral of motion (\ref{FibA}).

It is straightforward to show that for arbitrary $V_i$, the dynamical equations (\ref{Fib})
conserve a one-parameter family of quadratic invariants \addGF{(generalizations of the Manley-Rowe invariants for three-wave interactions)}:
\be{\cal F}_k=\sum\nolimits_{i=1} F_{i+k-1}|a_i|^2\,,\label{Fk}\ee where  $k$ could be of either sign if we define negative Fibonacci numbers: $F_{-j}=(-1)^{j+1}F_j$. All invariants can be obtained as linear combinations of any two of them. For example, the first two integrals are positive, independent, and in involution:
\be {\cal F}_1=\sum\nolimits_{i=1} F_i|a_i|^2\,,\quad {\cal F}_2=\sum\nolimits_{i=1} F_{i+1}|a_{i}|^2\ .\label{Fib1}\ee

In a closed system, the microcanonical equilibrium is ${\cal P}=\delta({\cal H}-C) \addGF{\delta({\cal F}_1-C_1)\delta({\cal F}_2-C_2)}$.
We now add dissipation and white-in-time pumping:
\be
\dot a_i=-\imath\partial {\cal H}/\partial a_i^*+\xi_i-\gamma_ia_i\ .\label{Hamb}
\ee
Here $\langle \xi_ia_j^*\rangle=\delta_{ij}P_i/2$.
\addGF{It is straightforward to show, also in a general case (\ref{HamW},\ref{Ham1}), that such forcing on average does not change the cubic Hamiltonian,  since $\langle \xi_ia_{i+1}a_{i+2}^*\rangle=P_i\langle \partial (a_{i+1}a_{i+2}^*)/\partial a_i^*\rangle=0$ for any $i$.} Denoting ${\cal H}_i=2\text{Re}  (V_ia_i^*a_{i-1}a_{i-2})$, we then obtain $\sum\nolimits_{i}d\langle {\cal H}_{i}\rangle/dt=-\sum\nolimits_{i} (\gamma_i+\gamma_{i-1}+\gamma_{i-2}) \langle {\cal H}_{i}\rangle$, which must be zero
in a steady state. At least when all sums $ \gamma_i+\gamma_{i-1}+\gamma_{i-2} $ are the same, ${\sum_i\langle  {\cal H}_i\rangle=}\langle { \cal H}\rangle=0$ (one can probably imagine exotic cases where separate $\langle {\cal H}_{i}\rangle\not=0$ but we shall not consider them).  If
pumping and damping are in a detailed balance, so that $\sum_k\alpha_kF_{i+k-1}= \gamma_i /P_i $ for every $i$, the thermal equilibrium distribution  is Gaussian: ${\cal P}=\exp(-\sum_k\alpha_k {\cal F}_k)$ --- it is a steady solution of the Fokker-Planck equation: 
\addGF{\bea \partial_t{\cal P} &=\{{\cal P},{\cal H}\}+\sum_i \Bigl[P_i\partial_{a_i}\partial_{a_i^*}+\gamma_i\bigl(\partial_{a_i}a_i
+\partial_{a_i^*}a_i^*\bigr)\Bigr]{\cal P}\nonumber\\&\propto \sum_i\left( 2  \gamma_i -P_i\sum_k\alpha_kF_{i+k}\right)=0\,.\nonumber\eea}
That solution realizes maximum entropy for given values of the invariants. The distribution is exactly Gaussian despite the system being described by a cubic Hamiltonian and thus  strongly interacting. The only restriction on the numbers $\alpha_k$ is normalization. In particular, when only $\alpha_1=1/\addGF{2}T$ is nonzero, we get the equilibrium equipartition with the occupation numbers $n_i\equiv\langle |a_i|^2\rangle=P_i/\addGF{2}\gamma_i=T/ F_i$.

In a turbulent cascade, the fluxes of the quadratic invariants can be expressed via the third cumulant. Gauge invariance and Zeckendorf theorem ensure that the triple cumulants are nonzero only for consecutive modes in the inertial range:
\begin{align}
J_i&\equiv \Im\langle a_i^*a_{i-1} a_{i-2}\rangle\,, \label{J} \\
F_{i+k-1}{d\langle |a_i|^2\rangle\over dt}&= 2F_{i+k-1}(V_{i-2}J_i-V_{i-1}J_{i+1}-V_{i}J_{i+2}) \nonumber
\\&=\Pi_k(i-1)-\Pi_k(i) =-\partial _i\Pi_k(i)\ .\label{Fib2}
\end{align}
The right hand side is the discrete divergence   of the flux \begin{align}
\Pi_k(m)&\equiv-\sum\limits_{i=1}^m F_{i+k-1}{d\langle |a_i|^2\rangle\over dt} \nonumber \\&= 2F_{m+k}V_{m-1}J_{m+1}+2F_{m+k-1}V_{m}J_{m+2}\ . \label{flux1}
\end{align}



The 3rd order  cumulants are zero in equilibrium, but in turbulence they are nonzero to carry the flux. In the inertial interval, the flux must be constant and its divergence zero. \addGF{For our class of models, we are able to find analytically the form of the 3rd cumulant (the analog of Kolmogorov's 4/5-law for fluid turbulence):
\be J_m= CF_{M-m+1}/ V_{m-2}\ ,\label{flux4}\ee
where real constant $C$ and integer $M$ can be of either sign. Let us substitute (\ref{flux4}) into (\ref{flux1}) and show that all the fluxes are non-zero constants independent of $m$:
\bea   &\Pi_k(m)=2F_{m+k}V_{m-1}CF_{M-m}/ V_{m-1}\nonumber\\&+2F_{m+k-1}V_{m}CF_{M-m-1}/ V_{m}=CF_{M+k-1}\ .\label{flux5}\eea}\addGF{The last  equality  follows from the Cassini identity}: $F_{m}F_{n}+F_{m-1}F_{n-1}=F_{m+n-1}$. All the fluxes have the same sign for any $k$, that is all the integrals ${\cal F}_k$ flow in the same direction \addGF{for such solutions}. 
We shall show in the next section what kind of fine-tuning is needed to get a double cascade when both cascades carry the same integrals.
In \cite{Lvo}, the (quadric) spectral flux of the (cubic) Hamiltonian was also defined, but  pumping does not produce it, so that $\langle{\cal H}\rangle=0$ in a steady turbulent state, as well as in thermal equilibrium.

\addGF{Every model of our family is completely characterized by specifying the dependence of $V_i$ on $i$. While thermal equilibrium does not depend on $V_i$ and is universal for the whole family, turbulence depends on $V_i$, as clear from  (\ref{flux4}). In what follows, we shall consider the power-law dependence $V_i=F_i^\alpha$, which turns into exponential dependence  $V_i\approx\phi^{i\alpha}$ for $i\gg1$. Therefore, the single real parameter $\alpha$ determines the model.   Our choice of particular values for $\alpha$ below will make the connection between wave and hydrodynamical turbulence through the Fibonacci model more explicit.}

\section{Cascade direction}\label{sec:direction}

To get an analytic insight into our turbulence, particularly, to understand the flux direction, consider an invariant sub-space of solutions with purely imaginary $a_k=i\rho_k$ for all $k$:
 \be {\partial\rho_i\over\partial t}=V_{i-2}\rho_{i-1} \rho_{i-2}-V_{i-1}\rho_{i-1}\rho_{i+1}-V_i\rho_{i+1}\rho_{i+2} \label{Fib4}\ee
In this case, 
${\cal H}\equiv0$. The invariant subspace owes its existence to the invariance of  (\ref{Fib}) with respect to the symmetry $a\to- a^*$.

Consider the chain running between some integers $M$ and $N$, either positive or negative, and assume $V_i/V_{i-1}=\phi^\alpha$. Then for  $\rho_i=A\phi^{i\beta}$ and $M+1<i<N-1$ we obtain:

\begin{equation}
\frac{\partial\rho_{i}}{\partial t}=A^{2}V_{i-2}\phi^{2i\beta}\left(\phi^{-3\beta}-\phi^{\alpha}-\phi^{2\alpha+3\beta}\right)\ .\label{Fib41}
\end{equation}
\begin{figure*}
  \includegraphics[width=17.5cm]{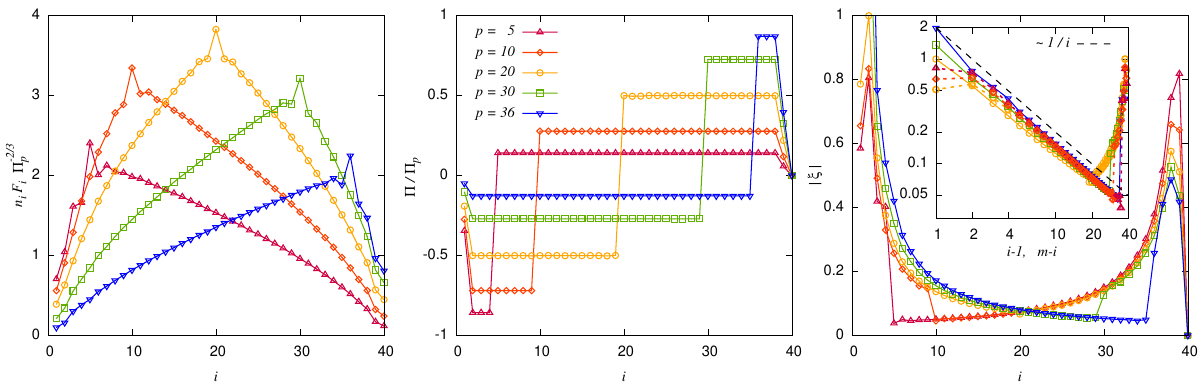} \centering
  \caption{
Compensated spectra, fluxes and skewness for $\alpha = 1/2$ with different pumping locations: $p=5$, 10, 20, 30, and 36 on the 40 mode interval.
Pumping rate is selected to provide the same flux in all cases, ${\Pi}_p = 67.65$.  In all cases damping rates are $\gamma_L = \gamma_R = 1$. Inset reproduces the longer arm of the cascades in log-log scale.
  }
\label{fig_sqrt_lever}
\end{figure*}

The right hand side of (\ref{Fib41}) turns into zero for $\beta=-(1+\alpha)/3$, which defines a steady solution $\rho_i=\phi^{-i(1+\alpha)/3}$ (also with the replacement $\phi\to-1/\phi$). This solution can describe either direct or inverse cascade, since the symmetry $\rho\to-\rho$, $t\to-t$ means that one reverses the flux by changing the sign of $\rho$ in this case.
Indeed,  consider the  evolution from the initial state where all amplitudes are zero except the first two $\rho_{M},\rho_{M+1}$. The first term in (\ref{Fib4}) then will produce $\rho_{M+2}$ of the same sign as $V_M\rho_M\rho_{M+1}$, which makes the flux positive, as it should be for a direct cascade.
Alternately, by pumping  the last two modes, the last term of (\ref{Fib4}) produces a negative flux.
Which cascade can be realized in reality: direct, inverse or both? Physically it is clear that the sign of the flux must be determined by the only parameter $\alpha$, that is by how mode interaction depends on the mode number. Indeed, for $\alpha=1/2$, the scaling of the flux steady solution coincides with that of the thermal equilibrium: $\langle\rho_i\rangle=0$, \addGF{$\langle \rho_i\rho_j\rangle=n_i\delta_{ij}=\delta_{ij}T/F_i\propto \phi^{-i}$, for $i\gg1$.}  Such state can be excited, for instance, by an imaginary pumping acting on every mode in detailed balance with dissipation. Physical common sense suggests that the cascade must carry the conserved quantity $\sum_iF_i\rho_i^2$ from excess to scarcity \cite{FF,ZLF}. For $\alpha>1/2$ the steady solution $\rho_i^2=\phi^{-2(1+\alpha)i/3}$ decays with $i$ faster than the equipartition $\rho_i^2\propto 1/F_i\propto \phi^{-i}$, so that it must correspond to a direct cascade. By the same token, we must have an inverse cascade for $\alpha<1/2$. \addGF{Of course, such consideration is a plausible argument, not a rigorous proof of the cascade sign.} Getting a little ahead of ourselves, mention here that we observe a double-cascade turbulence exactly at $\alpha=1/2$.

In a general complex case,  arguing that the cascade changes direction when $\alpha$ crosses $1/2$ is even less straightforward. The flux constancy determines the third moment, which only bounds the product of the second and fourth moments (the claim that it bounds the square root of the products of three second moments made in \cite{Pro1}  is incorrect). Yet a plausible argument can be made as follows. The input rate of ${\cal F}_k$ is equal to $\Pi=PF_{p+k-1}$ where $p$ is the position of the pumping. The input rate must be equal to the dissipation rate $\Pi= {2}\gamma_dF_{d+k-1}n_d$ for any choice of $\gamma_d$ taken at the dissipation position $d$. In order for $n_d$ to smoothly match the cascade, one must choose $\gamma_d$  comparable to the nonlinear interaction time: $\gamma_d\simeq V_dJ_d^{1/3}\simeq V_d(\Pi/V_dF_d)^{1/3}$. This gives an order-of-magnitude estimate $n_d\simeq (\Pi/V_dF_d)^{2/3}$. Such reasoning can be applied to every $i$, which in turn gives the estimate for the spectrum of occupation numbers: \be n_i\simeq(\Pi/V_iF_i)^{2/3}\,.\label{cascadeS}\ee  Since the direction of the flux is toward the occupation numbers that are lower than thermal equilibrium, $n_i\propto F_i^{-1}$, then again we see that the flux changes direction when $V_i\propto F_i^{1/2}$. The dimensionless degree of non-Gaussianity on such a spectrum,
\be\xi\equiv{J_i \over  n_i^{3/2}}\simeq{\Pi\over V_iF_i n_i^{3/2}}\simeq {PF_{p}\over V_iF_i n_i^{3/2}}\,,\label{nG}\ee must be  independent of $i$. For the spectrum close to equilibrium, $\xi\propto F_i^{3/2}/V_iF_i=F_i^{1/2}/V_i$.

\begin{figure}
  \includegraphics[width=8.5cm]{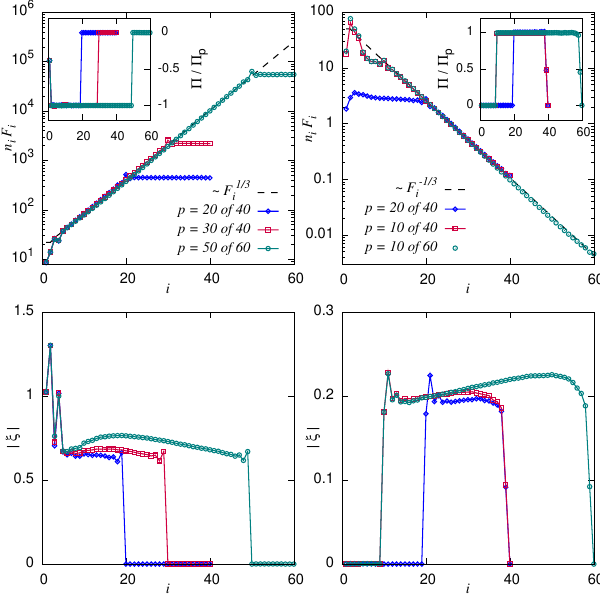} \centering
  \caption{
    Compensated spectra, fluxes, and the dimensionless skewness
    for $\alpha = 0$ (left) and $\alpha = 1$ (right)
    for systems with different location of pumping.
    In all cases $\Pi_p = 67.65$. For $\alpha = 0$ damping rates are $\gamma_L = 1.5$ and $\gamma_R =0$;
    for $\alpha = 1$ damping rates are $\gamma_L = 0$ and $\gamma_R = 140$ at mode 40 and $\gamma_R = 3500$ at mode 60.
  }
\label{spectra}
\end{figure}

Figures~\ref{fig_sqrt_lever}  and \ref{spectra} confirm these predictions.
We place the pumping at a single  mode, $i=p$, between two dissipation regions on the ends, letting the system to choose the cascade direction. The system~\eqref{Hamb} with pumping and damping has been evolved numerically using {\tt LSODE} solver \cite{lsode}.  At each step, random Gaussian noise of power $P$ is applied to the pumping-connected mode injecting flux $\Pi_p = P F_p$. Damping with $\gamma_L$  and $\gamma_R$  is applied to the two left-most and two right-most modes respectively.
For $\alpha=1/2$ ($V_i=\sqrt{F_i}$), the system is weakly distorted from equilibrium, with a constant flux on each
side of the pumping. For $\alpha \ne 1/2$ we find that the invariants are absorbed only on one end of the spectrum.
For $\alpha>1/2$ ($V_i=F_i$),  we have a thermal equilibrium 
to the left of pumping and the direct cascade (\ref{cascadeS}) with a constant $\xi$ to the right. In the opposite case  ($\alpha<1/2$, $V_i=$const), we find an inverse cascade  (\ref{cascadeS})  with constant $\xi$ to the left  and equilibrium equipartition  to the right of pumping.  In both cases, the damping on the flux side is carefully selected to avoid build-up in the spectrum (the damping on the equilibrium side can be then set to zero to establish cleaner scaling).
We have chosen  $V_i=F_i$ and $V_i=$~const because they qualitatively correspond to the Kolmogorov scaling of the direct energy cascade in incompressible turbulence and to the inverse wave action cascade in deep water turbulence respectively.

Thermal equilibrium at the scales exceeding the pumping scale \addGF{together with a direct cascade at smaller scales have} been predicted and observed \cite{Large}. \addGF{To the best of our knowledge}, nobody has seen before an inverse-only cascade \addGF{together with a thermal equilibrium on the other side of the pumping}, neither in hydrodynamic-type systems nor in wave turbulence or shell models.  {Inverse cascades play a prominent role in geophysics and astrophysics, from creation of planetary jets to Jupiter Great Red Spot and stormy seas.} In all known cases inverse cascades appear in systems with at least two conserved quantities that scale differently. All our conserved quantities (\ref{Fk}) scale the same in the limit $i\gg1$. \addGF{Probably  closest to our findings are the results of Tom and Ray \cite{TR} who observed an inverse cascade in the limiting case of a shell model with two invariants having the same scaling. Their inverse cascade had normal scaling and run from fast to slow modes; the direct cascade was not resolved, but was likely present.}

 Our observation poses the question:  can one find another class of systems with a single conservation law and the turbulent spectrum less steep than equilibrium. In weak wave turbulence, this requires the sum of the space dimensionality and the scaling exponent of the three-wave interaction to be less than the frequency scaling exponent \cite{ZLF}. We do not know such a physical system, nor we aware of any fundamental law that forbids its existence. Remark that the connection between the cascade direction, its stability and steepness relative to equipartition has been firmly established in the weak turbulence theory \cite{FF,ZLF}. In all known examples, the formal turbulent solution with a wrong flux sign is not realized; the system  chooses instead to stay close to equipartition with a slight deviation that provides for the flux in the right direction \cite{ZLF,FV1}.
Similarly, when we place pumping and damping at the ``wrong'' ends of a finite chain, our system heats up, staying close to thermal equilibrium. 

It is important that our system is a one-dimensional chain, as well as shell models, so that there is no space and consequently no distinction in the phase volume (number of modes) between infrared and ultraviolet parts of the spectrum. The directions along the chain are only distinguished temporally, i.e. in terms of growth/decay of the typical interaction time. The same combination $V_i^2/F_i\propto \phi^{2\alpha-1}$ determines the $i$-dependence of the inverse interaction time both for the equilibrium,  $V_ib_i^{1/2}=V_i F_i ^{-1/2}T^{1/2}$, and for a cascade, $V_i(\Pi/V_iF_i)^{1/3}=(V_i^2/F_i)^{1/3}\Pi^{1/3}$. As the above consideration shows, the cascade proceeds from slow modes to fast modes in Fibonacci turbulence. Similarly  in shell models \cite{Shell1,Shell2,Pro1} (albeit with parameters and conservation laws distinct from our model), a cascade  proceeding from fast modes to slow modes was never observed. It was argued that this is because the fast modes act like thermal noise on the slow ones, which must lead to equilibrium \cite{Shell1}. That this cannot be generally true follows from the existence of the inverse energy cascade in 2D incompressible turbulence and from numerous examples in weak wave turbulence where non-linear interaction time either grows or decays along the cascade. Moreover, the formation of the cascade spectrum proceeds from fast to slow modes (and not necessarily from pumping to damping), according to the information-theory argument \cite{SF}.

Why is the flux direction unambiguously related to the cascade acceleration in shell models in general and in our model in particular, in distinction from other cases? The argument can be made by considering capacity, a measure that tells at which end the conserved quantity is stored --- perturbations are known to run towards that end \cite{ZLF}. For example, the power-law energy density spectrum $\epsilon_k\propto k^{-s}$ in $d$ dimensions has the total energy $\int \epsilon_k\,d^dk$ --- at which end  it diverges is determined by the sign of $d-s$. This is generally unrelated to the direction of the energy cascade, determined by the sign of $s$, which tells whether the spectrum is more or less steep than the equipartition. However, in shell models the exponential character of $i$-dependencies makes the total energy $\sum_iF_i|a_i|^2$ determined by either the last or the first term of the sum, which solely depends on whether $ F_i|a_i|^2$ is steeper than equipartition or not, that is by the sign of the flux.

Which direction then the cascade goes in the symmetric case, $V_i=\sqrt{F_i}$? Now the naive  cascade solution  (\ref{cascadeS}) coincides with thermal equipartition, $F_in_i=$const, and the interaction time is independent of the mode number for such $n_i$. If we start from thermal equilibrium and apply pumping to some intermediate mode, the system develops cascades in both directions. The left panel of the Figure~\ref{fig_sqrt_lever} shows that the pumping at site $p$ inside the interval $(1,N)$ generates  left and right fluxes in the proportion $\Pi_L/\Pi_R\simeq (N-p)/p$. This seems natural as in the shorter interval the steeper spectrum falls away from the pumping, which must correspond to a larger flux. This means that if we want to keep the flux constant while increasing $p$ or $N-p$, we need to keep constant the ratio  $(N-p)/p$.

We end this section with a general remark. Fibonacci Hamiltonian is not symmetric with respect to reversing the order of modes, it sets the preferred direction, which is physically meaningful since the frequencies of two lower modes sum into the frequency of a high one. Yet, as we see in  the case $V_iF_i^{-1/2}=$const, direct and inverse cascades are pretty symmetric. So, it is natural to conclude that indeed the $i$-dependence of $V_iF_i^{-1/2}$ determines which way cascade goes.

\section{Along the cascades and away from equilibrium}

As we have seen, thermal equilibrium statistics is exactly Gaussian with no correlation between modes, despite strong interaction (which actually establishes equipartition). The reason for the absence of correlation is apparently the detailed balance that cancels them. We do not expect such cancelations in non-equilibrium states. In all cases of strong turbulence known before, the degree of non-Gaussianity  increases  along a direct cascade and stays constant along an inverse cascade \cite{Sym,Scol}. As we shall show now, non-Gaussianity always increases along the cascades in our one-dimensional chains.

\begin{figure}
  \includegraphics[width=8.5cm]{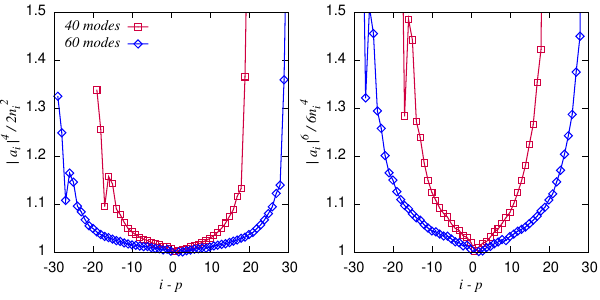} \centering
  \caption{
    Fourth and sixth moments for $\alpha = 1/2$ and center pumping in 40-mode system,
    with $\gamma_L = \gamma_R = 3$, $P=0.1$, and in 60-mode system with $\gamma_L=\gamma_R = 30$, $P=1$.
  }
\label{fig_sqrt_moments}
\end{figure}

We present first the symmetric case, where the system is close to the equilibrium equipartition with the temperature set by pumping and slowly changing with the mode number: $n_iF_i\approx (PF_p)^{2/3}f(i)$. The slow function $f(i)$ can be suggested by the analogy with the 2D enstrophy cascade \cite{FL,PF} as $f(i)\propto \ln^{2/3} F_i\propto i^{2/3}$, counting from the damping region. This gives the dimensionless cumulant $\xi\propto 1/i$. This hypothesis is supported by the right panel of the Figure~\ref{fig_sqrt_lever}, which shows that $\xi$ grows along both cascades by a power law in $i$ rather than exponentially. Let us stress that count always starts from the dissipation region, where we have the balance condition $\Pi=\gamma_dF_{d+k-1}n_d$ and where $\gamma_d\simeq V_dJ_d^{1/3}\simeq V_d(\Pi/V_dF_d)^{1/3}$ according to the dynamical estimate. This sets the nonlinearity parameter of order unity at the damping region and decaying towards pumping; the longer the interval, the smaller is $\xi$ at any fixed distance from the pumping region. The limit of long intervals may then be amenable to an analytical treatment. Indeed, Figure~\ref{fig_sqrt_moments} demonstrates that as the interval increases, the higher cumulants remain small over longer and and longer intervals starting from pumping. Despite the model having ultra-local interactions (every mode participates in only three adjacent interacting triplets), the cascade formation is very nonlocal. It is somewhat similar to thermal conduction: if we keep the flux but increase the distance, the distribution gets closer to the thermal equilibrium at every point.

\begin{figure}
  \includegraphics[width=8.5cm]{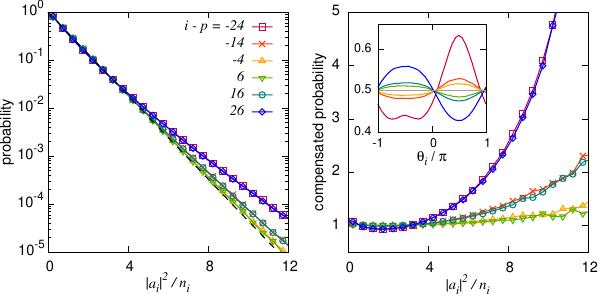} \centering
  \caption{
    Probability (left) and deviation of probability from equilibrium (right) for $\alpha = 1/2$.
    Main panels show probabilities of occupation numbers rescaled to their averages,
    the inset shows the probability
    of phase difference, $\theta_i = \varphi_i - \varphi_{i-1}  - \varphi_{i-2}$.
    Refer to the first panel for the line color for different modes.
    Data are shown for 60-mode system with center pumping and $\gamma_L=\gamma_R = 30$, $P=1$.
  }
\label{fig_sqrt_prob}
\end{figure}

\begin{figure}
  \includegraphics[width=8.5cm]{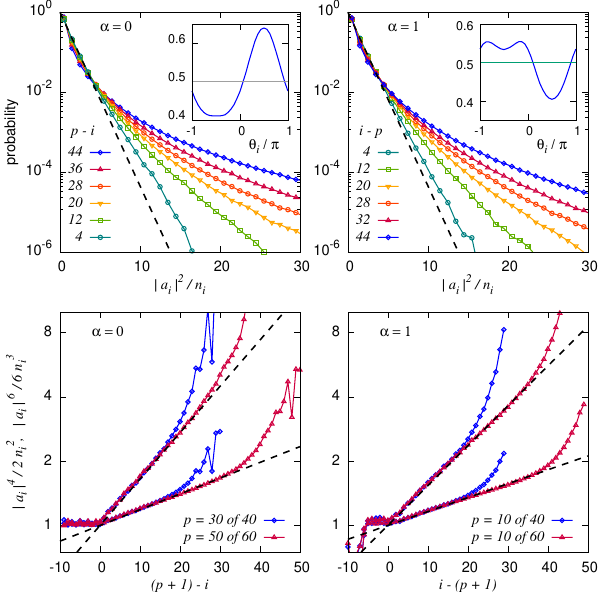} \centering
  \caption{Probabilities (top) and forth and sixth moments (bottom) for
    the inverse cascade, $\alpha = 0$ (left), and the direct cascade, $\alpha = 1$ (right).
    Probabilities for the rescaled occupation numbers are shown in the main panels,
    while probabilities for the phase difference,  $\theta_i = \varphi_i - \varphi_{i-1}  - \varphi_{i-2}$,
    are shown in the insets. The variation between ${\cal P}(\theta_i)$  for different $i$ is minor.
    In all cases $\Pi_p = 67.65$. For $\alpha = 0$, the damping rates are $\gamma_L = 1.5$ and $\gamma_R =0$;
    for $\alpha = 1$ the damping rates are $\gamma_L = 0$ and $\gamma_R = 140$ at $i=40$ and $\gamma_R = 3500$ at $i=60$. 
    \addGF{In the top panels the dashed lines indicate the Gaussian probability;
    in the bottom panels the dashed lines show linear fits to the data.}
  }
\label{prob}
\end{figure}

\begin{figure}
  \includegraphics[width=8.5cm]{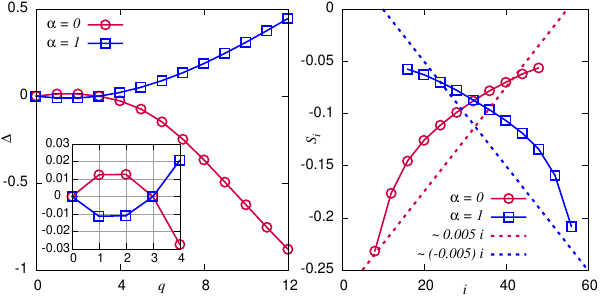} \centering
  \caption{\small Left panel: Anomalous exponents \addGF{computed as $\Delta(q)=q\zeta_3/3-\zeta_q$.
    Right panel: Decay of entropy down the cascade for the one-mode
    complex amplitude normalized by $\sqrt{n_i}$. The dashed lines  $S_i-S_p\approx -0.005|i-p|$ have the slopes
    equal  to $\Delta_1 \ln(\phi)$  with $\Delta_1$ shown in the left panel.} Direct cascade - blue, inverse cascade - red.}
\label{anom}
\end{figure}

Turning to asymmetric (one-cascade) cases, we see the cumulants higher than third growing with $F_i$ by a power law instead of logarithmic. Rather than look for scaling in the mode number $i$, we find it more natural to use $F_i$ (playing the role of frequency); at large $i$ one has  $F_i\approx \phi^i$, where $\phi$ is the golden mean. Traditional study of turbulence in general and shell models in particular was
focused on the single-mode moments (analog of structure functions), $\langle |a_i|^q\rangle \propto F_i^{-\zeta_q}$, whose anomalous scaling exponents, $\Delta(q)=q\zeta_3/3-\zeta_q$  give particular measures of how non-Gaussianity grows along the cascade. 
For $V_i=F_i^{\alpha}$, the flux law gives $J_i\propto \Pi/V_iF_i$, that is $\zeta_3=\alpha+1$.  The anomalous scaling is observable in numerics for the single-cascade cases  $\alpha=0$ and $\alpha=1$, as shown in the right panel of the Figure~\ref{anom}. This seems to be the first case of an anomalous scaling in an inverse cascade, with the anomalous dimensions having the opposite signs to those in direct cascades. The exponents start fairly small but grow fast with $q$.
The anomalous exponents, $\Delta(q)$,  can be related to the statistical Lagrangian conservation laws \cite{FGV,GA} in fluid turbulence; no comparable physical picture was developed for shell models. Without physical guiding, the set of the anomalous exponents is not very informative, all the more that they characterize only one-mode distribution.

Here we suggest a complementary set of three information-theoretic measures, which shed a new light on the turbulent statistics emerging along the cascade. The main distinction of any non-equilibrium state is that it has lower entropy than the thermal equilibrium at the same energy. Turbulence has the entropy that is much lower, which means that a lot of information is processed to excite the turbulence state. We pose the question:
where is the information that distinguishes turbulence from equilibrium encoded?


\vskip 0.5cm

\section{Where is the information encoded?}

First, the information is encoded in a single-mode statistics, which is getting more non-Gaussian deeper in the cascade. This must be reflected in the decay of the one-mode entropy, $S_i=S(x_i)=S(|a_i|/\sqrt{n_i})$, with the growth of $|i-p|$. This can be computed using the multi-fractal formalism: the moments  $\langle x_i^q\rangle \propto F_i^{-\zeta_q+q\zeta_2/2}$ in the limit of large  $|i-p|$ correspond to the multi-fractal distribution,
\be {\cal P}(x_i)\propto \int g(x_i/F_i^h)x_i^{-1}\exp[f(h)\ln F_i]\,dh\,,\label{LD1} \ee
where $f(h)=\min_q(\zeta_q-q\zeta_2/2-qh)$, that is $f(h)$ is the Legendre transform of $\zeta(q)$. The entropy is then $$S_i=-\int dx{\cal P}\ln {\cal P}\propto[\Delta'(0)-\Delta_2/2]\ln F_i\,.$$
This decay is logarithmic in frequency $F_i$, that is  linear in $i$, as indeed can be seen in Figure~\ref{anom}, where $i$ is counted from pumping.   
\addGF{Noticing that $\Delta_1 \approx \Delta_2$ and assuming quadratic dependence for $q \le 3$,
we estimate $\Delta'(0)\approx 3\Delta_1/2$ and observe that the dashed lines in the right panel of Figure~\ref{anom} with the slopes $\Delta_1 \ln \phi$  by the order of magnitude represent the entropy decay in the inertial interval in both direct and inverse cascades}.

\begin{figure}
  \includegraphics[width=8.5cm]{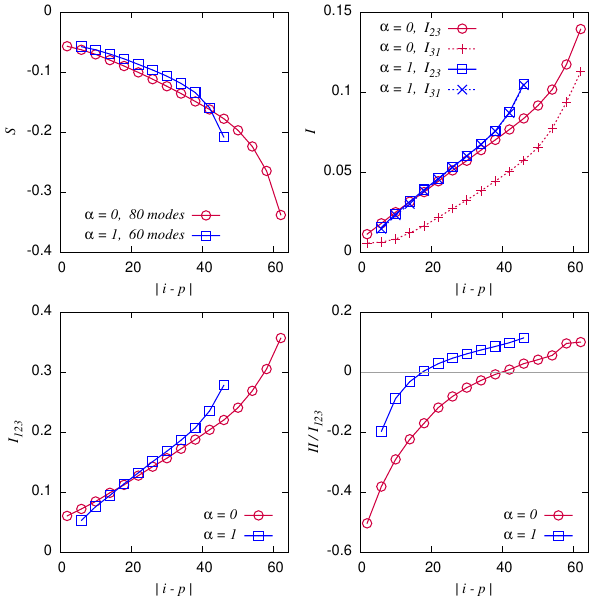} \centering
  \caption{
    Deviation of the entropy from equilibrium, the mutual information, and the interaction
    information (all in bits) for $\alpha = 0$  and $\alpha = 1$ and center pumping.
    Number of data points  $2\cdot10^8$ for $\alpha=0$, 80 modes,
    $6\cdot10^7$ for $\alpha = 0$, 60 modes, and $10^8$ for $\alpha=1$.
    For the bin size selected, all quantities agree with those obtained in a half-reduced data set.
  }
\label{mi}
\end{figure}

Second, the information is encoded in the correlations of different modes. It is natural to assume that correlations are strongest for modes in  interacting triplets, $a_i,a_{i+1},a_{i+2}$.
Disentangling of information encoded can be done  by using structured groupings \cite{MG,Bell,Eye}:
\bea & \sum_{i=1}^nS(a_i)-\sum_{ij}S(a_i,a_j)+\sum_{ijk}S(a_i,a_j,a_k)\\
&-\sum_{ijkl}S(a_i,a_j,a_k,a_l)+\ldots+(-1)^{n+1}S(a_1,\ldots,a_n)\nonumber\,.\label{COI}\eea
For $n=1$, this gives the one-mode entropy $S_i$ which measures the total amount of information one can obtain by measuring or computing one-mode statistics. While the entropy itself depends on the units or parametrization,  all the quantities (\ref{COI}) for $n>1$ are independent of units and invariant with respect to simultaneous re-parametrization of every single variable.  For $n=2$,  we have the widely used mutual information,
$$I_{ij}=S(a_i)+S(a_{j})-S(a_i,a_{j})\,,$$
which measures the amount of information one can learn about one mode by measuring another, that is characterizes the correlation between two modes. It is interesting that all pairs in the triplet have comparable mutual information in the direct cascade ($V_i=F_i$), while $I_{i,i+1}$ exceeds noticeably $I_{i,i+2}$ in the inverse cascade ($V_i=1$), see the upper right panel in Figure~\ref{mi}.
One can also define the total (multi-mode) mutual information as the relative entropy between the true joint distribution and the product distribution: $I(a_1,\ldots,a_k)= \sum_{i=1}^k S(a_i)-S(a_1,\ldots,a_k)$. It is positive and monotonically decreases upon averaging over any of its arguments.  As we see from  Figure~\ref{mi}, the changes along the cascade in one-mode entropy and in two-mode and three-mode mutual information are comparable, that is one obtains comparable amount of information about turbulence from these quantities.

\begin{figure}
  \includegraphics[width=8.5cm]{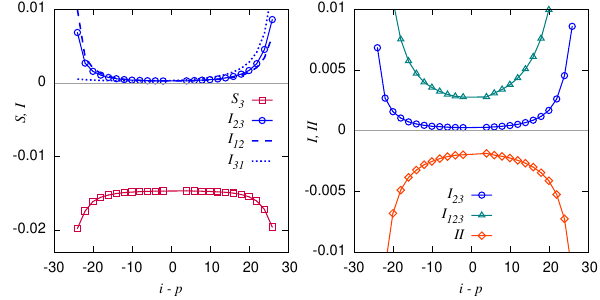} \centering
  \caption{
    Deviation of entropies from equilibrium, mutual information, and interaction
    information for $\alpha = 1/2$ and center pumping
    for a set of $5 \cdot 10^7$ data point.
    The same values of entropy were obtained for  a set of $2 \cdot 10^7$ data point, that is $S_i$ is saturated.
    Both $I$ and $II$ show a slight decrease in absolute values with the increase of the ensemble size from  $2 \cdot 10^7$ to $5 \cdot 10^7$.
  }
\label{mi_v_eq_F}
\end{figure}

To see how much more information one gets by measuring or computing the three modes simultaneously compared to separately by pairs, one needs to use the measure of the irreducible information encoded in triplets, as given by the third member of the hierarchy (\ref{COI}):
\begin{align}
II_{i}=&S(a_i)+S(a_{i+1})+S(a_{i+2})+S(a_i,a_{i+1},a_{i+2})\nonumber\\
&-S(a_i,a_{i+1})-S(a_i,a_{i+2})-S(a_{i+1},a_{i+2})
\nonumber\\
=&I_{i,i+1}+I_{i,i+2}+I_{i+1,i+2}-I_{i,i+1,i+2}\nonumber\\
=&I(i,i+1)-I(i,i+1|i+2)\,,\label{II}
\end{align}
It is called interaction information in the classical statistics and topological entanglement entropy in the quantum statistics \cite{MG,KP}. Interaction information measures the influence of the third  variable  on the amount of information shared between the other two and could be of either sign.
Positive $II(X,Y,Z)$ measures the redundancy in the information about $Y$ obtained  by measuring  $X$ and $Z$ separately, while negative one measures synergy which is the extra information about $Y$ received by knowing $X$ and $Z$ together. While we cannot prove it mathematically, it seems physically plausible that systems with three-mode interaction must demonstrate synergy. Indeed, one finds a strong synergy in weak turbulence: it was shown that $I_{123}\gg I_{12}+I_{23}+I_{13}$ \cite{SF}, so that $II<0$ and much more information is encoded in three modes than in the pairs separately. Here we find that the same is true for the cascades close to thermal equilibrium at $V_i=\sqrt{F_i}$ as seen in Figure~\ref{mi_v_eq_F}. Indeed, the two-mode mutual information is much smaller than both the one-mode entropy and the absolute value of the interaction information, which is negative.

Let us stress that both the mutual information and the interaction information are symmetric, that is they measure the degree of correlation rather than causal relationship or cascade direction.

We compute the entropies and  mutual information as follows. First, we obtain the probability distribution in 4D space ($x_{i-2}^2, x_{i-1}^2, x_i^2, \theta_i)$ and integrate it to get corresponding 1D and 2D distributions.  Here,  $\theta_i = \varphi_i - \varphi_{i-1} - \varphi_{i-2}$ , where $\varphi_i$ is the phase of mode $i$, and $x_i=|a_i|/\sqrt{n_i}$, while  $n_i=\langle |a_i|^2\rangle$ is the direct average.  Mutual information and information interaction are computed directly from entropies, $S = -\Sigma {\cal P} \log_2 {\cal P}$, obtained for these distributions, since all normalization factors cancel out in subtraction. The entropy for an individual mode, however, is presented relative to the Gaussian entropy based on the average occupation number obtained for the binned, staircase distribution for $x_i^2$.  We use the
bin sizes $\Delta x^2_i =1$ for $\alpha=0$ and $\alpha=1$, and  $\Delta x^2_i = 1/2$ for $\alpha = 1/2$. In  all cases  $\Delta \theta = 2\pi/32$.

Far from equilibrium, we find synergy for the modes close to the pumping and redundancy for damping, see the last panel of Figure~\ref{mi}. That means that the interaction information passes through zero in the inertial interval. There even seems to be a tendency to stick to zero in the inertial interval but this requires further studies with the number of modes exceeding our present abilities. (Our computations are done with a record number of modes, up to 80, while previous studies were mostly done for 20-30. The interaction times decrease exponentially with the mode number, which imposes heavy requirements on the computational time step. On top of that one needs very
long runs to collect enough statistics to reliably represent the three-mode probability distribution in four-dimensional space.) With the present set of data we can suggest that  most of  the information about the three-mode correlation is in the sum of the pair correlations in the triplet. This is more pronounced in the direct cascade than in the inverse cascade.
Since the requirements on statistics grow exponentially with the dimensionality, the suggestion that  one can get most
of information (or at least a large part of it) from lower-dimensional probability distributions is great news for turbulence measurements and modeling. To put it simply, comparable amounts of information can be brought from one-mode and from three-mode measurements in direct and inverse cascades; most of that information can be inferred from two-mode measurements. It remains to be seen to what degree  this property of small (asymptotically zero?) interaction information is a universal feature of strong turbulence.

Insets in the Figures~\ref{fig_sqrt_prob},\ref{prob} show the probability distribution of the relative phase, $\theta_i$,
which is closely related to the flux (skewness),  proportional to $\langle |a_ia_{i-1}a_{i-2}|\sin\theta_i\rangle$. The probability maximum is then at $\pm\pi/2$ for direct and inverse cascades respectively. Also, the $i$-dependence of the phase distributions is in accordance with the changes in skewness along $i$. In the two-cascade symmetric case, the distribution is flat (the phases are random) near the pumping, and the phase correlations appear along the cascades, as can be seen comparing the last panel of Figure~\ref{fig_sqrt_lever} with the inset in the right panel of Figure~\ref{fig_sqrt_prob}. In the one-cascade cases, both skewness and the form of the spectrum are practically independent of the mode number, as seen from  Figures~\ref{spectra},\ref{prob}.

The fact that the deviations from Gaussianity grow along our inverse cascade, in distinction from all the inverse cascades known before, calls for reflection. We used to think about the anomalous scaling and intermittency in spatial terms: Direct cascades proceed inside the force correlation radius, which imposes non-locality, while in inverse cascades one effectively averages over many small-scale fluctuations, which bring scale invariance \cite{Sym,Scol}. The emphasis on the spatial features was reinforced by the success of the Kraichnan's model of passive tracer turbulence, where it has been shown that  the spatial (rather than temporal) structure of the velocity field is responsible for an anomalous scaling and intermittency of the tracer. There is no space in our case, so apparently it is all about time. Indeed, as we have seen, all our cascades propagate from slow to fast modes, which leads to the build-up of non-Gaussianity and correlations. As a result, the entropy of every mode decreases and the inter-mode information grows along the cascade. This diminishes the overall entropy compared to the entropy of the same number of modes in thermal equilibrium with the same total energy.

Despite qualitative similarity, there is a quantitative differences between our direct and inverse cascades.
Figures~\ref{prob},\ref{anom}  show that the one-mode statistics and its moments faster deviate from Gaussian as one proceeds along the inverse cascade than the direct one. And yet one can see from  Figures~\ref{anom},\ref{mi_v_eq_F} that the one-mode entropy is essentially the same in both cascades, as well as the mutual information between two neighboring modes and the three-mode mutual information. The mutual information between non-neighboring modes $I_{13}$ is about twice smaller, as seen in Figure~\ref{mi}. This difference can probably be related to the dynamics, which in our system is the coalescence of two neighboring modes into the next one and the inverse process of decay of one into two. In the dynamical equation (\ref{Fib4}),  only one (first) term is responsible for the direct process (and the direct cascade), while  two  terms are responsible for the inverse process (and the inverse cascade).

An important distinction between double-cascade and single-cascade turbulence in our system is the dependence on the system size. The degree of non-Gaussianity of the complex amplitudes is fixed in the dissipation regions of the double cascade, so that in the thermodynamic limit the statistics is Gaussian in the inertial intervals. On the contrary, the statistics of the amplitudes is fixed at the forcing scale for a single cascade, and it deviates more and more from Gaussianity as one goes along the cascade.

{We end this section} by a short remark on the production balance of the total entropy $S=-\langle\ln\rho(a_1,\ldots,a_N)\rangle$. Here $\rho(a_1,\ldots,a_N)$ is the full $N$-mode PDF. Since wave interaction does not change the total entropy, then the entropy absorption by the dissipation must be equal to the entropy production by the pumping \cite{GFF,SF}:
\be P  \int\! \prod_i \frac{da_{i}da_{i}^{*}}{{2  \rho}} \Bigl|{\partial\rho\over\partial a_p}\Bigr|^2=  2\sum_k\gamma_k ,
.\label{stop1}\ee
For a single-cascade cases ($V_i=1$ and $V_i=F_i$), the energy balance $PF_p= {2}\gamma F_dn_d$ means that the left hand side of (\ref{stop1}) must be much larger than the Gaussian estimate $P/n_p$  \cite{SF}. It may seem to contradict our numerical finding that the pumping-connected mode $a_p$ has its one-mode statistics close to Gaussian. Of course, there are nonzero triple correlation and the mutual information with two neighboring modes in the direction of the cascade. Yet since $\xi\simeq 1$, then the triple moment $J_p\simeq n_p^{3/2}$  both in direct and inverse cascades, so that the contribution to the left hand side of (\ref{stop1}) is comparable with $P/n_p$. We conclude then that even the pumping-connected mode must have strong correlations with many other modes. Since the triple correlation function of non-adjacent modes are zero, such correlations must be encoded in higher cumulants. That deserves further study.

\section{Kolmogorov multipliers and self-similarity}

\begin{figure}
  \includegraphics[width=8.5cm]{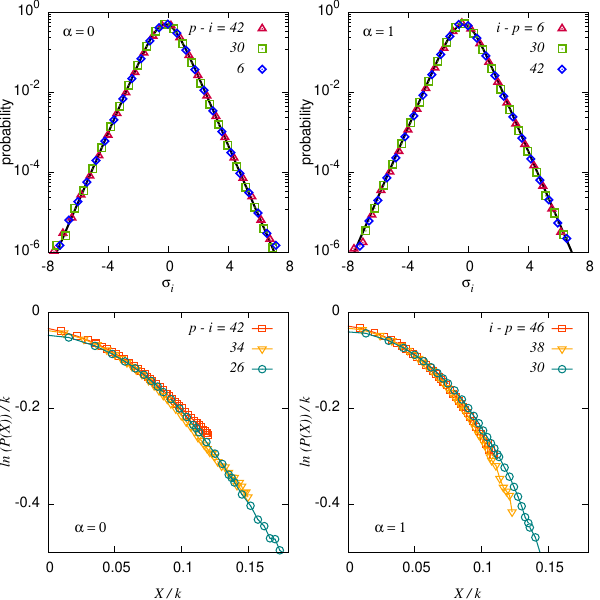} \centering
  \caption{Top: probability distributions of the Kolmogorov multipliers $\sigma_i=\ln|a_i/a_{i-1}|$ for different positions in the inverse (left) and direct (right) turbulent cascades. Solid lines correspond to the thermal equilibrium  $P(\sigma)=1/2\cosh^2(\sigma-\bar\sigma)$, where $\bar\sigma=-(1/3)\ln\phi$ for the inverse cascade and  $\bar\sigma=-(2/3)\ln\phi$ for the direct one. Bottom: probability distributions of \addGF{$X=\ln|a_k|^2$} collapse to the large-deviation form far away from the pumping, that is for large $k=|i-p|$.
  }
\label{fig_Kmult.pdf}
\end{figure}

Unbounded decrease of entropy along a single cascade prompts one to ask whether the total entropy of turbulence is extensive (that is proportional to the number of modes) or grows slower than linear with the number of modes, so
there could be some ``area law of turbulence" (like for the entropy of black holes). This  question can be answered with the help of the so-called Kolmogorov multipliers, $\sigma_i=\ln|a_i/a_{i-1}|$ \cite{Kol}. Figure~\ref{fig_Kmult.pdf} shows that in our cascades the multipliers have universal statistics independent of $i$, similar to shell models \cite{multi,Kol1,Kol2,Mail}. One consequence of the scale invariance of the statistics of the multipliers is that the entropy of the system is extensive, that is proportional to the number of modes. Of course, the entropy depends on the representation. From the information theory viewpoint, the Kolmogorov multipliers  realize representation by (almost) independent component, that is allow for maximal entropy. In other words, computing or measuring turbulence in terms of multipliers gives maximal information per measurement (the absolute maximum is achieved by using the flat distribution, that is the variable $u(\sigma)$ defined by $du=P(\sigma)d\sigma$).

The amplitudes are expressed via the multipliers:  $$X_k=\ln x_k =\ln{|a_k|\over\sqrt{n_k}} {=\ln x_p +\sum_{i=p+1}^{p+k}\sigma_i+\frac{1}{2}\log\frac{n_{p}}{n_{k}}}\,.$$
The first term is due to the pumping-connected mode, which correlates weakly with $\sigma_i$ in the inertial interval. As shown below, the correlation between multipliers decays fast with the distance between them. That suggests that the   statistics of the amplitude logarithm at large $k$ must have asymptotically a large-deviation form:
\be \ln{\cal P}(X_k)=-kH(X_k/k)\ .\label{LD2}\ee Indeed, the three upper curves in the top row of Figure~\ref{prob} collapse in these variables, as shown in the bottom row of Figure~\ref{fig_Kmult.pdf}. The self-similar distribution of the logarithm of amplitude, (\ref{LD2}), is a dramatic simplification in comparison with the general multi-fractal form (\ref{LD1}). Technically, it means that  $g(x_k/F_k^h)=g(e^{X_k-kh\ln\phi})$ is such a sharp function that the integral in  (\ref{LD1}) is determined by the single $X_k$-dependent value, $h(X_k)=X_k/k\ln\phi$. We then identify $f =-H/\ln \phi$.

The self-similarity of the amplitude distribution (plus the independence of the phase distribution on the mode number) is great news, since it allows one to predict the statistics of long cascades  (at higher Reynolds number) from the study of shorter ones. In our case, Figure~\ref{fig_Kmult.pdf} shows that 28-th mode already has the form close to asymptotic.  Self-similarity and finite correlation radius of the Kolmogorov multipliers has been also established experimentally for Navier-Stokes turbulence \cite{Kol3}. To avoid misunderstanding, let us stress that the self-similarity is found for the probability distribution of the logarithm of the amplitude, which does not contradict the anomalous scaling of the amplitude moments with the exponents $\zeta_q$ determined by the Legendre transform of $f$ or $H$.

If the multipliers were statistically independent, one would compute  $\ln{\cal P}(X)=-kH(X/k)$ or $\zeta_q$ proceeding from  $P(\sigma)$ by a standard large-deviation formalism: $H(y)=\min_z[zy-G(z)]$, where $G(z)=\ln\int d\sigma  e^{z\sigma}P(\sigma)$. Such derivation would express  $\langle |a_k|^q\rangle$ via  $\langle e^{q\sigma_k}\rangle$, which is impossible since the former moments exist for all $q$, while the latter do not because of the exponential tails of $P(\sigma)$, see also \cite{Kol3,Kol4}.

Therefore, to describe properly the  scaling of the amplitudes one needs to study correlations between multipliers. Physically, it is quite natural that the law of the distribution change along the cascade must be encoded in correlations between the steps of the cascade. Indeed, we find that the neighboring multipliers are dependent, albeit weakly, as expressed in their mutual information (traditionally used pair correlation function \cite{Kol1,Kol2,Kol3} is not a proper measure of correlation for non-Gaussian statistics). We find that for the inverse cascade,
$I(\sigma_i,\sigma_{i+1})\simeq 0.23$,  $II(\sigma_i,\sigma_{i+1},\sigma_{i+2})\simeq-0.1$. For the direct cascade, 
$I(\sigma_i,\sigma_{i+1})\simeq 0.3$, $II((\sigma_i,\sigma_{i+1},\sigma_{i+2})\simeq -0.08$. No discernible $I(\sigma_i,\sigma_{i+k})$ were found for $k>1$. While $\sigma_i$ and $\sigma_{i+2}$ are practically uncorrelated, there is some small synergy in a triplet.

To appreciate these numbers, let us present for comparison the statistics of the Kolmogorov multipliers in thermal equilibrium. Normalized for zero mean and unit variance, we have
\begin{align}
&P(\sigma)=\int\int_0^\infty\!\! dxdy\,e^{-x-y}\delta\bigl(\sigma-{1\over2}\ln{x\over y}\bigr)={1\over2\cosh^2\sigma}, \nonumber\\
&P(\sigma_i,\sigma_{i+1})={8e^{4\sigma_i+2\sigma_{i+1}}\over
	\bigr[1+e^{2\sigma_i}\bigl(1+e^{2\sigma_{i+1}}\bigr)\bigr]^{3}}\ .\label{pss}
\end{align}


That gives $I(\sigma_i,\sigma_{i+1}) =\ln 2-1/2\approx 0.19$.

\begin{figure}
  \includegraphics[width=8.5cm]{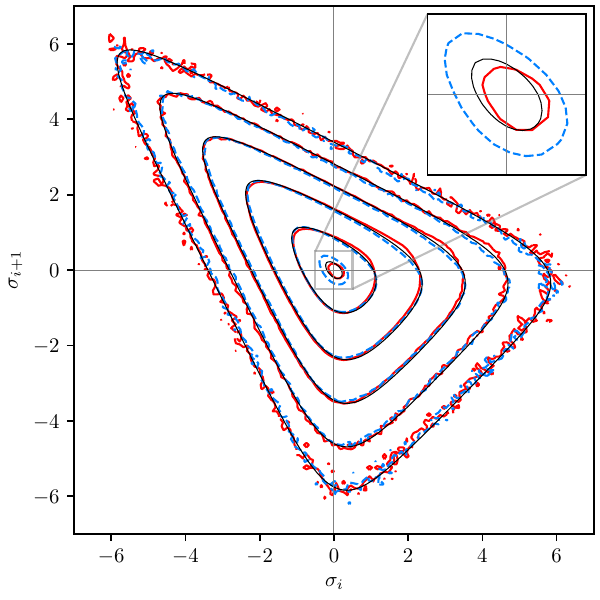} \centering
  \caption{Joint probability distributions of two neighboring Kolmogorov multipliers shifted to zero means. The contours are at $log_{10}(P) = -0.55, -1, -2, -3,  -4, -5$.  Inverse cascade ($\alpha=0$) is red, direct cascade ($\alpha=1$) is blue, black is the equilibrium distribution (\ref{pss}).
  }
\label{fig_Kmult2D.pdf}
\end{figure}

Figure~\ref{fig_Kmult.pdf} shows that the equilibrium Gaussian statistics of independent amplitudes perfectly  represents the statistics of a single multiplier.
The joint PDFs $P(\sigma_i,\sigma_{i+1})$ are shown in Figure~\ref{fig_Kmult2D.pdf} for thermal equilibrium and for two cascades. Again, the Gaussian statistics represents turbulence remarkable well. The differences between the three cases are most pronounced around the peak at the origin, while the distant contours are hardly distinguishable. In plain words, the probabilities of strong fluctuations of the multipliers are the same in thermal equilibrium as in turbulence cascades.   This is remarkably different from the statistics of the complex amplitudes, which demonstrate most difference between the three cases for strong fluctuations and for high moments. There seems to be a certain duality between fluctuations of the amplitudes and multipliers: strong fluctuations of the multipliers correspond to weakly correlated amplitudes, while strong fluctuations of the amplitudes may require their strong correlations and thus correspond to multipliers close to their mean values. Whether this duality can be exploited for an analytic treatment remains to be seen.  The information about the anomalous scaling exponents of the amplitudes in turbulence must be encoded in the correlations between multipliers. Note that the mutual information $I(\sigma_i,\sigma_{i+1})$ for both cascades ($I=0.23$ and $I=0.30$) is not that much higher than in thermal equilibrium ($I=0.19$ bits). Physicists tend to be much excited about any broken symmetry; it is refreshing to notice that relatively small information is needed to encode the broken scale invariance in turbulence. How to decode this information from  the joint statistics of multipliers remains the task for the future

\section{Discussion}

 {The most surprising finding of our work} is the existence of an inverse-only cascade and its anomalous scaling. In all cases known before, an inverse cascade appears only as  {an outlet} for an extra invariant that cannot  {be transferred} along the direct cascade with other invariant(s). In  a truly weak turbulence, when the whole statistics is close to Gaussian, an inverse-only cascade is indeed impossible, since it would require an environment that provides rather than extracts entropy, which contradicts the second law of thermodynamics \cite{GFF,SF}.  Here we have shown that an inverse-only cascade is possible in a strong turbulence. As far as an anomalous scaling is concerned, we relate it to the change of the interaction time along the cascade.
All the inverse cascades known before  run from fast to slow modes and have a normal scaling. In our case, as in all shell models, cascades always proceed from slow to fast modes. Apparently, this is the reason that non-Gaussianity increases along all our cascades, and an anomalous scaling takes place in both single inverse and single direct cascades. Indeed, proceeding from fast to slow modes (in inverse cascades known before) involves an effective averaging over fast degrees of freedom, which diminishes intermittency. On the contrary, our cascades build up intermittency as they proceed.

Another unexpected conclusion follows from the entropy production balance in a steady turbulent state: even though the marginal statistics of the pumping-connected mode (averaged over all other modes) can be close to Gaussian, the correlations of that mode with other modes cannot be weak.

Most of the present work was devoted to disentangling of the information encoded in strong turbulence. It was predicted that in weak turbulence most of the information is encoded in the three-mode statistics \cite{SF}, and  {Figure~\ref{mi_v_eq_F}} confirms this prediction. Yet in strong turbulence, we find that as much information is encoded in one-mode as in two-mode statistics, while three-mode statistics does not add much. This  {could be of practical importance for turbulence studies since it is much more difficult to collect, store and analyze statistics for three-mode and multi-mode distributions.} Another important lesson is that measuring or computing mode amplitudes (or velocity structure functions) brings diminishing returns, that is less and less information, as one goes deep into the cascade. The maximal information is encoded in the statistics of the Kolmogorov multipliers. Most of that information is encoded in the statistics of a single multiplier; less than 10\% is encoded in the correlation of neighbors. How to decode it is the task for the future.

We wish to thank Yotam Shapira for helpful discussions.  The work was supported by the Scientific Excellence Center and Ariane de Rothschild Women Doctoral Program at WIS, grant  662962 of the Simons  foundation, grant 075-15-2019-1893 by the Russian Ministry of Science,  grant 873028 of the EU Horizon 2020 programme, and grants of ISF, BSF and Minerva. NV was in part supported by NSF grant number DMS-1814619. This work used the Extreme Science and Engineering Discovery Environment (XSEDE), which is supported by NSF grant number ACI-1548562, allocation DMS-140028.

%


\begin{thebibliography}{1}

\bibitem{Obukhov} A. M. Obukhov, \emph{Integral invariants in hydrodynamic systems}, Dokl. Akad. Nauk SSSR, 184:2, (1969).
	
\bibitem{Peierls} R. Peierls, \emph{On the kinetic theory of thermal conduction in crystals}, Ann. Phys. (N.Y.) \textbf{3}, 1055 (1929).

\bibitem{ZLF} V. Zakharov, V. Lvov and G. Falkovich, \emph{Kolmogorov Spectra of Turbulence} (Springer 1991).

\bibitem{NR} A. Newell and B. Rumpf, \emph{Wave turbulence}, Annu. Rev. Fluid Mech. \textbf{43}, 59 (2011).

\bibitem{Naz} S. Nazarenko, {\it Wave Turbulence} (Springer 2011).

\bibitem{Kart} E. Kartashova, L. Piterbarg and G. Reznik, \emph{Weak Nonlinear-interactions of rossby waves on sphere}, Oceanology {\bf 29}, 405 (1990).

\bibitem{Bif} L. Biferale,v, \emph{Shell models of energy cascade in turbulence}, Ann. Rev. Fluid Mech {\bf35} :441-468 (2003).

\bibitem{Lvo} V. L'vov E. Podivilov and I. Procaccia, \emph{Hamiltonian structure of the Sabra shell model of turbulence: exact calculation of an anomalous scaling exponent}, EPL \textbf{46} 609 (1999).

\bibitem{Pro} V. L'vov, E. Podivilov, A. Pomyalov, I. Procaccia
and D. Vandembroucq, \emph{Improved shell model of turbulencetext},Phys. Rev. E \textbf{58} 1811,22 (1998).

\bibitem{FF} J.D. Fournier and U. Frisch, \emph{d-Dimensional turbulence}, Phys Rev A {\bf17}, 747 (1978).

\bibitem{Pro1}T. Gilbert, V. L'vov, A. Pomyalov and I. Procaccia,\emph{Inverse cascade regime in shell models of two-dimensional turbulence}, Phys. Rev. Lett. \textbf{89}, 074501 (2002).


\bibitem{lsode} A.C. Hindmarsh, ACM Signum Newsletter, \textbf{15}(4), 10 (1980),  K. Radhakrishnan and A. C. Hindmarsh, the Livermore solver for ordinary differential equations (1993).

\bibitem{Large} E. Balkovsky, G. Falkovich, V. Lebedev and I.Y. Shapiro, \emph{Large-scale properties of wave turbulence},
Phys. Rev. E {\bf52} 4537 (1995).

\bibitem{TR} R. Tom and S. Ray, \emph{Revisiting the SABRA model: Statics and dynamics},  Eur.Phys.Lett, \textbf{120} (2017)

\bibitem{FV1}G. Falkovich and N. Vladimirova, \emph{Cascades in nonlocal turbulence}, Phys. Rev. E \textbf{91}, 041201(R) (2015).

\bibitem{Shell1}E. Aurell, G. Boffetta, A. Crisanti, P. Frick, G. Paladin, and A. Vulpiani, \emph{Statistical mechanics of shell models for two-dimensional turbulence}, Phys. Rev. E \textbf{50}, 4705 (1994).

\bibitem{Shell2} P. D. Ditlevsen and I. A. Mogensen, \emph{Cascades and statistical equilibrium in shell models of turbulence}, Phys. Rev. E \textbf{53}, 4785 (1996).


\bibitem{SF} M. Shavit and G. Falkovich, \emph{Singular measures and information capacity of turbulent cascades}, Phys. Rev. Lett. \textbf{125}, 104501 (2020).

\bibitem{Sym} G. Falkovich, \emph{Symmetries of the turbulent state}, J. Phys. A: Math. Theor. \textbf{42} 123001 (2009).

\bibitem{Scol} G. Falkovich, \emph{Cascade and scaling}, Scholarpedia, 3(8) 6088 (2008).

\bibitem{FL} G.Falkovich and V. Lebedev, \emph{Nonlocal vorticity cascade in two dimensions}, Phys Rev E \textbf{49}, R1800 (1994);  Phys Rev E \textbf{50} 3883 (1994).


\bibitem{PF} C. Pasquero and G. Falkovich, \emph{Stationary spectrum of vorticity cascade in two-dimensional turbulence}, Phys Rev E \textbf{65}, 056305 (2002).

\bibitem{FGV} G. Falkovich, K. Gawedzki and M. Vergassola, \emph{Particles and fields in fluid turbulence}, Rev. Mod. Phys. (2001).

\bibitem{GA} G. Falkovich and A. Frishman, \emph{Single flow snapshot reveals the future and the past of pairs of particles in turbulence}, Phys. Rev.  Lett. \textbf{110}, 214502 (2013).

\bibitem{MG} W.J McGill, \emph{Multivariate mutual information}, Psychometrika. \textbf{19},  97-116 (1954).

\bibitem{Bell} A.J. Bell, in International workshop on independent component analysis and blind signal separation (2003) p. 921.

\bibitem{Eye} Cluster decomposition of the information encoded in different subsets of interacting particles was found in unpublished notes of Onsager (Gregory Eyink, private communication).

\bibitem{KP} A. Kitaev and J. Preskill, \emph{Topological entanglement entropy},  Phys. Rev. Lett. \textbf{96}, 110404 (2006).

\bibitem{GFF} G. Falkovich and A. Fouxon, \emph{Entropy production and extraction in dynamical systems and turbulence}, New. J. Phys. \textbf{6}, 50 (2004).

\bibitem{Kol} A. Kolmogorov, \emph{A refinement of previous hypotheses concerning the local structure of turbulence in a viscous incompressible fluid at high Reynolds number}, J. Fluid Mech. \textbf{13} 1 (1962).

\bibitem{multi}R. Benzi, L. Biferale and G. Parisi, \emph{A random process for the construction of multiaffine fields}, Physica D \textbf{65} (1993).

\bibitem{Kol1}G. Eyink, S. Chen, and Q. Chen, \emph{The joint cascade of energy and helicity in three-dimensional turbulence}, J. Stat. Phys.
\textbf{113}, 719 (2003).

\bibitem{Kol2} L. Biferale, A. A. Mailybaev, and G. Parisi, \emph{Optimal subgrid scheme for shell models of turbulence}, Phys. Rev. E
\textbf{95} 043108 (2017).

\bibitem{Mail} A.A. Mailybaev, \emph{Hidden scale invariance of intermittent turbulence in a shell model}, Phys. Rev. Fluids \textbf{6.1} (2021).


\bibitem{Kol3} Q. Chen, S. Chen, G. L. Eyink, and K. R. Sreenivasan, \emph{Kolmogorov’s third hypothesis and turbulent sign statistics},
Phys. Rev. lett. \textbf{90} 254501 (2003).

\bibitem{Kol4} G.  Eyink, S. Chen, Q. Chen, \emph{Gibbsian Hypothesis in Turbulence}, J. Stat. Phys. \textbf{113} (2003).




















\end{thebibliography}
\end{document}